\documentclass[runningheads]{llncs}

\usepackage[height=235mm,width=155mm,center]{crop}

\usepackage{graphicx}
\usepackage{diagbox}
\usepackage{makecell}
\usepackage{float}
\usepackage{colortbl}
\usepackage{subfig}
\usepackage[flushleft]{threeparttable}
\usepackage{paralist}

\usepackage{booktabs}

\usepackage[export]{adjustbox}

\newcommand{\censor}[1]{\textit{\scriptsize\textless withheld during blind review\textgreater}}
\newcommand{\censorshort}[1]{\textit{\scriptsize\textless withheld\textgreater}}
\newcommand{\censorinvisible}[1]{}
\newcommand{\censorimage}[2]{#2}

\renewcommand{\censor}[1]{#1}
\renewcommand{\censorshort}[1]{#1}
\renewcommand{\censorinvisible}[1]{#1}
\renewcommand{\censorimage}[2]{#1}

\newlength{\Oldarrayrulewidth}
\newcommand{\Cline}[2]{%
	\noalign{\global\setlength{\Oldarrayrulewidth}{\arrayrulewidth}}%
	\noalign{\global\setlength{\arrayrulewidth}{#1}}\cline{#2}%
	\noalign{\global\setlength{\arrayrulewidth}{\Oldarrayrulewidth}}}

\begin{document}

\title{Is This Really You? An Empirical Study on Risk-Based Authentication Applied in the Wild}
\titlerunning{Is This Really You? An Empirical Study on RBA Applied in the Wild}

\author{Stephan Wiefling\inst{1}%
	 \and
	Luigi {Lo Iacono}\inst{1}%
	\and
	Markus D\"urmuth\inst{2}}

\institute{TH K\"oln - University of Applied Sciences, Cologne, Germany\\
	\email{\{stephan.wiefling, luigi.lo\_iacono\}@th-koeln.de} \and
    Ruhr University Bochum, Germany
	\email{markus.duermuth@rub.de}
}
\authorrunning{S. Wiefling et al.}

\maketitle

\newcommand\blfootnote[1]{%
	\begingroup
	\renewcommand\thefootnote{}\footnote{#1}%
	\addtocounter{footnote}{-1}%
	\endgroup
}
\blfootnote{\scriptsize Postprint version of a paper accepted for IFIP SEC 2019. The final publication is available at Springer via %
	 \url{http://dx.doi.org/10.1007/978-3-030-22312-0\_10}}

\vspace{-1.5em}
\begin{abstract}
	Risk-based authentication (RBA) is an adaptive security measure to strengthen password-based authentication. RBA monitors additional implicit features during password entry such as device or geolocation information, and requests additional authentication factors if a certain risk level is detected. RBA is recommended by the NIST digital identity guidelines, is used by several large online services, and offers protection against security risks such as password database leaks, credential stuffing, insecure passwords and large-scale guessing attacks. %
	Despite its relevance, the procedures used by RBA-instrumented online services are currently not disclosed. Consequently, there is little scientific research about RBA, slowing down progress and deeper understanding, making it harder for end users to understand the security provided by the services they use and trust, and hindering the widespread adoption of RBA.	
    
	In this paper, with a series of studies on eight popular online services, we
	\begin{inparaenum}[(i)]
		\item analyze which features and combinations/classifiers are used and are useful in practical instances,
		\item develop a framework and a methodology to measure RBA in the wild, and
		\item survey and discuss the differences in the user interface for RBA.
	\end{inparaenum}
	Following this, our work provides a first deeper understanding of practical RBA deployments and helps fostering further research in this direction.
\end{abstract}

\vspace{-1.5em}
\section{Introduction}
Weaknesses in password-based authentication have been known for a long time \cite{morris_password_1979}. They range from weak and easy to guess passwords \cite{bonneau_science_2012,wang_targeted_2016} or password re-use~\cite{das_tangled_2014} to being susceptible to phishing attacks. Still, passwords are the predominant authentication mechanism deployed by online services today~\cite{bonneau_passwords_2015,quermann_state_2018}. To increase the users' security, service operators should implement additional measures. \emph{Two-factor authentication (2FA)}~\cite{petsas_two-factor_2015} is one widely offered measure that improves account security, but is rather unpopular (e.g. in January 2018, less than 10~\% of active Google accounts used 2FA~\cite{milka_anatomy_2018}). \emph{Risk-based authentication (RBA)}~\cite{freeman_who_2016} is an approach that increases security with minimal impact on user interaction, and thus has the potential to provide secure authentication with good usability. It is among the approaches suggested by the NIST digital identity guidelines to mitigate online guessing attacks~\cite{grassi_digital_2017}.

\subsubsection{Risk-based Authentication (RBA)}
\label{section:rba}
RBA is typically used in addition to passwords or other forms of user authentication. It is designed to protect against a rather strong attacker that either knows the correct credentials (i.e., username / password pair) or can guess correct credentials with a low number of guesses. Examples include \emph{credential stuffing attacks}~\cite{wang_polymorphism_2014} where an attacker tries credentials leaked from another service, \emph{phishing attackers}, or \emph{online guessing attacks}~\cite{wang_targeted_2016}.
During password entry RBA monitors and records additional features that are contextually available. In principle, a number of various distinct features can be taken into account (see Table~\ref{tab:rba-feature-comparison}), including  the \emph{IP address} and derived features such as \emph{geolocation} or \emph{country}, and the \emph{user agent}~\cite{bonneau_privacy_2014,freeman_who_2016}.
Some features are better suited for risk assessment than others: The IP address, e.g., could be rated as ``more important'' than the user agent string since spoofing an IP address is considered as more difficult than the latter~\cite{alaca_device_2016}.

From these features a \emph{risk score} is calculated. %
It is then typically classified into three buckets (low, medium and high risk)~\cite{freeman_who_2016,molloy_risk-based_2012,hurkala_architecture_2014}%
. Depending on the risk score and its classification, a variety of actions can be performed by the service. When a risk score exceeds e.g. the low threshold and falls into the medium risk category, the service typically requests additional authentication factors from the user (e.g. verification of email address or phone number~\cite{iaroshevych_improving_2017,shepard_using_2014,freeman_who_2016}), requires to solve a CAPTCHA~\cite{shepard_using_2014}, or informs the user about suspicious activities~\cite{google_notifying_2016}. If the risk score is deemed high, the service can decide to block access altogether, but this event is rare, as it will not allow legitimate users mistakenly classified as a high risk to recover. The thresholds of when a user becomes suspicious have to be carefully chosen for each individual RBA use case scenario%
.

\subsubsection{Contribution}
\label{section:contribution}
We investigate how RBA is used on eight high-traffic online services (Amazon, Facebook, GOG.com, Google, iCloud, LinkedIn, Steam and Twitch). We created 28 virtual identities and 224 user accounts for this purpose. During a period of 3.5 months we conducted studies to determine (an approximation to) a set of features that contributes to the risk score computation, and studied the influence of these features. We also captured and analyzed the deployed additional authentication factors%
. Our studies revealed serious vulnerabilities emphasizing the need for an open discussion on RBA in science.

To achieve reliable and repeatable results, we developed an automated browser testing framework and simulated human-like user behavior with individual activities on each of the online services. The framework contains enhanced technical camouflage measures to be indistinguishable from human users. The developed testing framework\footnote{Provided as open source software at \url{https://github.com/DASCologne/HOSIT}} can be used to analyze black boxed services for RBA features.
Our work is intended to support both research and development. Researchers benefit from an increased transparency on the current practice of RBA deployment. Also, they obtain a test methodology and tooling for running replication or follow-up studies. Developers obtain guided insights on how to best create or improve own RBA implementations. The same is true for administrators aiming at integrating RBA as an additional line of defense in their online services. This all contributes to an open scientific discussion on RBA, ultimately leading to a comprehensively understood security measure, leaving no room for obscurities. We hope that public research on RBA will enable a broader adoption of RBA and thus protect a larger user base, while currently only larger online services are capable to offer RBA techniques (beyond very basic and inaccurate service).

\subsubsection{Outline}
The rest of the paper is organized as follows. Section~\ref{section:related_work} reviews related work. Section~\ref{section:system-and-method} describes the developed automated testing framework, created identities and prerequisites for the studies. The study setup and obtained results are described in Section~\ref{section:experiments}. We discuss findings and limitations in Section~\ref{section:discussion} and conclude with the main contributions and an outlook on future work in Section~\ref{section:conclusion}.

\section{Related Work}
\label{section:related_work}
The features and authentication factors deployed by RBA-instrumented online services are currently either not disclosed or just briefly mentioned by the respective companies~\cite{milka_anatomy_2018,iaroshevych_improving_2017,johansson_risk-based_2014}. This lack hinders any scientific debate and rigorous analysis to facilitate the effective and open use of RBA. These debates and analyses are even more important today since RBA is recommended by NIST~\cite{grassi_digital_2017} and therefore becoming a requirement for many IT security professionals.

Most of the RBA-related research is focused on evaluating the reliability and robustness of certain features. A RBA method based on mouse and keyboard dynamics was developed and tested by Traore et al. \cite{traore_combining_2012}. Judging from the observed equal error rate, they concluded that this method is not suitable for RBA inside the login process. %
Hurkala and Hurkala \cite{hurkala_architecture_2014} published a software architecture of a RBA system. The features \textit{IP address}, \textit{login time}, \textit{availability of cookie}, \textit{device profiling} and \textit{failed login attempts} are implemented in the RBA system. The limitations and effectiveness of these features were not estimated. Freeman et al. \cite{freeman_who_2016} presented the, to the best of our knowledge, first publicly known RBA algorithm using \textit{IP address} and \textit{user agent} as features. Steinegger et al. \cite{steinegger_risk-based_2016} presented another RBA implementation, with \textit{browser fingerprint}, \textit{failed login attempts} and \textit{IP based geolocation} as features. Alaca and van Oorschot \cite{alaca_device_2016} classified and rated 29 distinct methods for device fingerprinting regarding possible \textit{``distinguishing info''}. They rated \textit{IP address} and \textit{geolocation} as \textit{``high''}. Daud et al. \cite{daud_adaptive_2017} introduced an adaptive authentication method applying HTML5 canvas fingerprinting. The effectiveness of this method is unclear due to the lack of testing with participants. Herley and Schechter \cite{herley_distinguishing_2018} presented a method for authentication servers to distinguish attacks from legitimate traffic. They rated the \textit{password used for a failed attempt} as a strong feature to identify attacks.

Petsas et al. \cite{petsas_two-factor_2015} estimated the quantity of Google user accounts with enabled 2FA functionality. They used headless browser automation with enhancements for user simulation. Their methodology, using browser automation and observing reactions, is roughly similar to ours. However, due to the complex nature of RBA and novel browser automation detection methods~\cite{vastel_detecting_2018}, a considerably higher amount of effort was necessary in our studies.%

\section{Black Box Testing RBA}
\label{section:system-and-method}

In this section we introduce the developed approach for black box testing RBA implementations in the wild. The basic methodology is to create accounts on the inspected online services and to observe the behavior when accessing the service using these accounts for a variety of scenarios. This seemingly simple procedure is complicated by a number of factors:
\begin{inparaenum}[(i)]
	\item~The account's login history may influence the risk score. Thus, testing multiple scenarios with the same account may produce unreliable results.
	\item~Automated testing is likely influencing the outcome, as one of the tasks of RBA is specifically to detect bots.
	\item~The list of features that potentially may be used by online services to determine the risk score is vast, and simply testing all combinations is next to impossible.
	\item~Depending on the service's implementation of RBA, the feedback can be coarse-grained, i.e., giving mostly binary information (RBA triggered/not triggered), while other online services provide more fine-grained information.
\end{inparaenum}

Our approach considers these issues and mitigates their effects on the results. %
We created a larger number of virtual identities and spent several weeks to train them on legitimate behavior. The data collection uses an extensively patched version of Chromium and a careful planning to protect against detection.

\subsection{Creation of Identities}
\label{subsection:identities}
We created 28 identities for our studies. %
User accounts for all eight inspected online services were created with each identity. We used a random identity generator for identity creation. Each identity consisted of first and last name, birthday, gender (50\%~male, 50\%~female), job title (function, company) as well as typing speed. Each identity owns an individual IP address (geolocation: \censorshort{TH K\"oln}) and a personal computer (virtual machine running Ubuntu Linux 16.04 LTS).
We conducted a one month pilot phase with one identity in order to optimize our identity creation, training and testing automation%
. Afterwards, we started the automated training and testing with the remaining 27 identities.
The account creation for Facebook required some extra care, as RBA is not activated per se for all accounts~\cite{iaroshevych_improving_2017}. We manually conducted extra training to these accounts (e.g. friend requests) prior to the studies. Resulting of the higher effort, 14 Facebook accounts (5 male, 9 female) were created. Six accounts (4 male, 2 female) were suspended during training because of ``suspicious'' activities. Since female accounts had higher success rates in terms of accepted friend requests or messages, we preferred them in Facebook account creation.
Thus, in total we created 224 accounts of which 210 remained available for training and 204 for inspecting the targeted online services.

\subsection{Training of Identities}

\label{subsubsection:browsing-behavior}
Each online service was trained with individual user activities for each identity in a 3.5 month period between December 2017 and March 2018. %
Each identity executed 20 user sessions lasting between 1.5 and 2 hours within a training period of 2 or 4 weeks. The start of the browsing sessions varied randomly between two time spans (9:00 - 9:30 AM, 1:00 - 1:30 PM) to mitigate possible automation detection by the online services. For further mitigation, the identities were created iteratively in small batches of three to four identities per week.

We developed individual automated user activities for each online service. Activities include the login process, actions on the online services at logged in state (\textit{user action}) and the logout process. %
In the login process, our user opens the targeted online service in a new browser tab, enters its login credentials and accesses this service. We considered typical user activities for the user actions, e.g. scrolling in the news feed or browsing on the online service. These actions included randomness and fine-grained variations to avoid being spotted as a ``scripted human''. Also, the user behavior differed between genders. %
For the logout, our user logs out of the online service and closes the tab.

We simulated browsing activities on other websites in separate tabs, as online services may track this browsing behavior~\cite{bujlow_survey_2017}. Users visited a search engine and looked for current events in local media. They followed some of the links in the search results and ``read'' the website's content by scrolling and waiting.

These activities were conducted inside browsing sessions. Each session was initiated with an empty browser history including cookies and local cache. The cookies were retained inside each browsing session. Afterwards, the testing sequence of online services was shuffled to a random order. We did this to prevent that our user logs into online services at the same time throughout the study.

\subsection{Implementation of RBA Inspection System}
\label{subsection:camouflage-enhancements}
The implemented RBA inspection system is based on the browser Chromium 64.0.3253.3. For browsing automation, the library Puppeteer 0.13.0 is used. The obtained observations during the test phase are stored in a MongoDB log.

Chromium was operated in a custom \emph{headful mode} (browser is launched with visible graphical user interface inside a virtual window session). We used the headful mode to avoid detection of our automated browsing. When Chromium is executed in \emph{headless mode}, which is specifically designed for browsing automation, a number of differences in Chromium's behavior allow websites to detect the automation mode~\cite{vastel_detecting_2018}. In fact, during pilot testing we experienced situations in which inspected online services treated a browser in headless mode differently.

Furthermore, we modified the Chromium source code to minimize possible detection of our automated RBA inspection system. %

We implemented the user automation framework using Puppeteer, a library to control Chromium. We found that several of the provided automation functions can be detected by online services. The constant delay in the standard Puppeteer key typing function is used to detect automated input. %
We therefore modified and enhanced several Puppeteer library functions to mimic human behavior more closely:
\begin{inparaenum}[(i)]
	\item~We added randomized delays between pressing and releasing key buttons as well as consecutive button presses.
	\item~We adjusted the default mouse input behavior of clicking on the exact center of a specified element by selecting a random click point in the center quarter of the element. Moreover, the default time between pressing and releasing the mouse button of zero was replaced with a more realistic randomized click time.
    \item~We implemented a scrolling function to imitate human-like reading of website contents.
\end{inparaenum}

We integrated external services providing CAPTCHA solving capabilities in order to allow our RBA inspection system to operate fully automated. 

\subsection{Inspection of RBA Features}
\label{subsection:rba-features}

\begin{table}[t]
			\caption{Comparison of possible RBA features (bold: selected  for the studies)}
		\medskip
		\centering
		\resizebox{0.85\linewidth}{!}{
			\begin{threeparttable}
				\begin{tabular}{@{}lll@{}}
				    \toprule
					\textbf{Feature} & \textbf{RBA references }                         & \textbf{Distinguishing} \\
					& \quad \textbf{(except \cite{alaca_device_2016})} & \quad \textbf{info} \cite{alaca_device_2016} \\
					\midrule
					\textbf{IP address}$^\#$ & \cite{freeman_who_2016,golan_system_2013,cser_forrester_2012,akhtar_real_2011,hurkala_architecture_2014,steinegger_risk-based_2016} & High \\
					\textbf{User agent string} & \cite{freeman_who_2016,hurkala_architecture_2014,spooren_mobile_2015} & High* \\
					\textbf{Language} & \cite{freeman_who_2016,hurkala_architecture_2014,cser_forrester_2012} & High* \\
					\textbf{Display resolution} & \cite{daud_adaptive_2017,spooren_mobile_2015} & High* \\
					\textbf{Login time} & \cite{freeman_who_2016,hurkala_architecture_2014,golan_system_2013,spooren_mobile_2015,cser_forrester_2012} & Low$^+$ \\
					\hline
					Evercookies & \cite{hurkala_architecture_2014} & Very high \\
					Canvas fingerprinting & \cite{daud_adaptive_2017,molloy_risk-based_2012,steinegger_risk-based_2016} & Medium \\
					Mouse and keystroke dynamics & \cite{traore_combining_2012} & - (\textit{Low for scroll wheel fingerprinting})\\
					Failed login attempts & \cite{hurkala_architecture_2014,steinegger_risk-based_2016} & - \\
					WebRTC & - & Medium \\
					Counting hosts behind NAT & - & Low \\
					Ad blocker detection & - & Very low \\
					\bottomrule
				\end{tabular}
				\begin{tablenotes}
					\small
					\item $^\#$ Includes IP based \textit{geolocation}.
					\item * Refers to \textit{major software and hardware details}
					\item $^+$ Refers to \textit{system time and clock drift}. Alaca and van Oorschot did not consider the login time. Hurkala and Hurkala \cite{hurkala_architecture_2014} estimated a medium risk level for unusual login times.
				\end{tablenotes}
			\end{threeparttable}
		}
		\label{tab:rba-feature-comparison}
\end{table}

A wide variety of features can be used for RBA deployments, ranging from browser provided information to network information~\cite{traore_combining_2012,alaca_device_2016}. To reduce complexity, we selected five features based on the number of mentions in literature and the evaluations in~\cite{alaca_device_2016} in terms of highest \textit{``distinguishing info''} (see Table~\ref{tab:rba-feature-comparison}).

We selected the features \textit{IP address}, \textit{user agent string}, \textit{language}, \textit{login time} and \textit{display resolution} for our investigations.

Canvas fingerprinting and evercookies provide a high level of information~\cite{alaca_device_2016,daud_adaptive_2017,acar_web_2014}. Canvas fingerprinting can be seen as a more robust and fine-grained version of user agent strings. Evercookies can uniquely identify a device. Since both features are considered as harder to fake, they add a high level of trust, possibly bypassing RBA security mechanisms. Since we aimed to test the ``uncertain'' area in terms of RBA risk scores, we did not consider both for our studies.

Prior to the study design, we estimated possible risk score results for specific variations of feature values. We used these estimations to design the final studies. Since no public information on the analyzed RBA implementations was known, we considered three publications \cite{freeman_who_2016,hurkala_architecture_2014,cser_forrester_2012} as a baseline for the estimation. We made use of the maximum possible range of ratings. However, since IP addresses are considered as more spoofing resistant than the other features~\cite{alaca_device_2016}, we expect this feature to be weighted highest inside the black box RBA implementations.

We assume that the \textit{IP address} risk score increases with both geographical distance towards the usual values and changes in IP address and internet service provider (ISP). Since users are more likely moving in their current region, we expect the risk score to be \textit{medium} at a maximum inside the same country. We assume changes in continents to be more unusual, so we expect a \textit{high} risk score in that case. We rated the risk score for IP addresses of the anonymization service \textit{Tor} as \textit{unknown} for two reasons:
\begin{inparaenum}[(i)]
  \item~Tor exit nodes (and Tor users) can be identified through a public list. Thus, one publication~\cite{hurkala_architecture_2014} estimated a high risk score for Tor.
  \item~Facebook explicitly supports Tor. Hence, lower risk scores can also be possible.
\end{inparaenum}
We subdivided the \textit{user agent string} into \textit{browser}, \textit{operating system} (OS) and \textit{version}. We expect users to switch browsers more likely than the OS, which is why we weighted browser changes lower than those in the OS. For the remaining three features, we assume that changes in one or more parameters will increase the score equally.

\section{Studies}
\label{section:experiments}
In this section, we describe the setup and results of the studies we conducted to evaluate the eight analyzed online services for their RBA behavior. We conducted two studies. In the first one, we tested how the online services reacted to six different variations of IP addresses to reduce the number of required test conditions for the second study (see Section~\ref{subsection:study-1}). In the second and main study we then determined which of the five investigated features (see Table~\ref{tab:rba-feature-comparison}) play a role in RBA decision-making (see Section~\ref{subsection:study-2}). We tested all possible combinations of these features for each online service and observed the results. We did this to determine whether a certain feature was included in the online service's feature set and to ascertain how a particular feature was weighted in the online service's RBA decision-making. Finally, we also did several activities on user accounts so that online services might offer diverse selections of additional authentication factors. We did this to capture as many additional authentication factors applied by the targeted online services as possible (see Section~\ref{subsection:study-3}). An extended version of our results including all captured dialogs can be found online~\cite{wiefling_rbainfo_2019}.

\subsection{Study 1: Determining IP Feature Thresholds}
\label{subsection:study-1}
The feature space that can be used for RBA is huge, and even with the restrictions put forth in Section~\ref{subsection:rba-features} the search space is still too large for the type of study we envision. Even the particularly important IP address feature has a wide range of possible values. Possibly interesting variations range from dynamic IPs (same ISP, same geolocation) or different access points (work, home, mobile) at similar locations, to national or international travelling or Tor (see Table~\ref{tab:versuch11}). Thus, in a first step we treated the IP space separately and tried to find thresholds for the individual online services that are close to the decision boundary of the decision procedure. This will simplify the subsequent experiments and reduce the number of required probes.

\begin{table}[b]
	\vspace{-1.5em}
	\centering
	\caption{Setup of study 1 to determine the RBA triggering threshold for the IP feature}
	\resizebox{0.8\linewidth}{!}{
		\begin{tabular}{l || l l l l}
		    \toprule
			& \textbf{IP}    & \textbf{ISP}                     & \textbf{Geolocation}                    & \textbf{Description}\\
			\midrule
			probe 0 & fixed & \censorshort{TH K\"oln} & \censorshort{Cologne, Germany} & same IP as used during training\\
			\midrule
			probe 1 & fresh & \censorshort{TH K\"oln} & \censorshort{Cologne, Germany} & fresh IP in the same building \\
			probe 2 & fresh & \censorshort{Netcologne}& \censorshort{Cologne, Germany} & different provider in the same city\\
			probe 3 & fresh & AWS                     & Frankfurt, Germany             & same country, different provider\\
			probe 4 & fresh & AWS                     & Paris, France                  & same continent, different provider\\
			probe 5 & fresh & AWS                     & Oregon, USA                    & different continent\\
			\midrule
			probe 6 & fresh & random                  & random (Tor exit node)         & Tor exit node at random location \\
			\bottomrule
		\end{tabular}
	}
	\label{tab:versuch11}
	\vspace{-0.3cm}
\end{table}

\subsubsection{Methodology}  %
\label{subsection:empiricalderivation}
In this first study varied the IP address only. We equipped seven of the trained identities with new IP addresses (Table~\ref{tab:versuch11}). Probe 0 uses the identical IP from which the online services were trained before. Probe 1 and probe 2 are located in close vicinity of the training IP (same city, physical distance less than 1~km), where probe 1 is from the same ISP (a university) and probe 2 is from a different ISP. Probes 3 to 5 used IPs with an increasing distance from the training origin. We used VPN tunnels through Amazon Web Services (AWS) instances for these probes. Probe 6 used the Tor network, with an IP of the exit node that is potentially known by service providers and sometimes treated differently%
. Logins at all online services were conducted with the new IP address and reactions of the online services were recorded.

\begin{table}[t]
	\centering
	\caption{Results of study 1 showing the determined RBA triggering thresholds for the IP feature (bold lines).} 
	\resizebox{\linewidth}{!}{ 
		\begin{threeparttable}
			\centering
			\begin{tabular}{l l ||c c c c c c c c} 
				\textbf{IP variation} & \textbf{Identity} & \textbf{Facebook} & \textbf{Google} & \textbf{Amazon} & \textbf{LinkedIn} & \textbf{GOG.com} & \textbf{Steam} & \textbf{Twitch} & \textbf{iCloud} \\ 
				\hline
				probe 0 (\censorshort{TH K\"oln}, fixed) & \textit{All identities} & - & - & - & - & - & - & - & - \\
				\hline
				\Cline{2pt}{7-7}
				probe 1 (\censorshort{TH K\"oln}, fresh) & IDA, IDAA$^+$ & - & - & - & - & A & - & - & - \\
				probe 2 (\censorimage{Netcologne}{same city, different ISP}) & IDB & - & S & - & - & A & - & \O & - \\ 
				probe 3 (Frankfurt) & IDC & - & S & - & - & A & - & - & - \\ 
				\Cline{2pt}{4-6}
				probe 4 (Paris) & IDD & - & A & A & A & A & - & - & - \\ 
				probe 5 (Oregon) & IDE & - & A & A  & A & A & - & - & - \\ 
				\hline
				probe 6 (Tor) & IDF & - & A & A & A & A & O & - & - \\ 
				\Cline{2pt}{3-3} \Cline{2pt}{8-10}
			\end{tabular}
			\vspace{0.5em}
			\begin{tablenotes}
				\small
				\item A: Additional authentication factors requested \hspace{1em} O: CAPTCHA displayed before login
				\item S: Security alert submitted (via email) \hspace{4.5em} \O: reCAPTCHA not displayed before login
				\item - : No RBA triggered \hspace{12.2em} $^+$: Facebook login was conducted with this identity
			\end{tablenotes}
		\end{threeparttable}
	}
	\label{tab:ergebnisseVersuch1Teil1}
\end{table}

\subsubsection{Results}
The obtained results are depicted in Table \ref{tab:ergebnisseVersuch1Teil1}. We see that the thresholds seem to be at IP variation probe 4 (Google, Amazon, LinkedIn) and probe 1 (GOG.com). Facebook, Steam, Twitch and iCloud did not request additional authentication factors, if only the IP address was varied. A CAPTCHA inside the Steam login form was visible in probe 6 (Tor). A reCAPTCHA on the Twitch login form was not displayed in probe 2\censorinvisible{ (Netcologne)} while being visible vice versa. These might rather be signs for blacklisting (Steam) or whitelisting (Twitch) than for RBA. Google sent an email containing a security alert on two occasions before reaching the threshold of asking for additional authentication factors.

Based on the results, we extracted three IP settings for use in the subsequent experiments. These were selected for each online service separately, reflecting the determined thresholds. We set probe~0 \censorinvisible{(TH K\"oln)} for GOG.com, probe~3 (Frankfurt) for Google, Amazon and LinkedIn as well as probe~5 (Oregon) for Facebook, Steam, Twitch and iCloud. We did not use Tor in subsequent studies, due to its unpredictable nature (frequent variations of IP addresses) which could produce unreliable results. Varying the ISP to AWS (probe~3) inside the same country did not result in requesting additional authentication factors. Hence, we assume that using AWS IP addresses did not affect the reliability of our results.

\subsection{Study 2: Examining RBA Usage}
\label{subsection:study-2}
In the second and main study, we determined which features play a role in the overall RBA decision-making and under which circumstances the inspected online services request additional authentication factors.

\subsubsection{Methodology}
We tested all 31 possible combinations of the five parameters \textit{IP address}, \textit{user agent string}, \textit{language}, \textit{time parameters} and \textit{display resolution} for triggering RBA. Each trained account conducted one or two login attempts with different parameter combinations. The \textit{IP address} was chosen one step beneath the determined RBA triggering threshold. The remaining parameters were chosen to represent the highest possible risk estimation as defined in Section \ref{subsection:rba-features} (see Table \ref{tab:testparameters}). We chose a far distance country with a different national language than in the training country as the testing country. Based on the online services' behavior of all 31 parameter combinations, we are able to derive possible feature set parameters.

\begin{table}[b]
  \vspace{-2em}
  \caption{Setup of study 2 showing the probed features. We tested all possible combinations, i.e., $2^{5}-1=31$ variations per online service.
  }
  \centering
  	\resizebox{0.8\linewidth}{!}{
	  \begin{threeparttable}
		    \begin{tabular}{@{}lll@{}}
		        \toprule[1pt]
		                   & \textbf{Neutral/Training} & \textbf{Testing} \\
		      \midrule
		      \textbf{IP address} & as in training & as determined in Sect~\ref{subsection:study-1} \\
		      \textbf{User agent} & Chrome/Linux & Firefox/Windows 10 \\
		      \textbf{Languages} & \censorshort{de-DE,de,en-US,en} & es-MX,es,en-US,en \\
		      \textbf{Time} \quad  \textbf{Timezone} & UTC+1 \censorinvisible{(Europe/Berlin)} & UTC-6 (Mexico/General) \\
		         \hspace*{11mm}    \textbf{Login times [UTC+1]} & 9:00 AM - 2:30 PM & 0:00 - 1:00 AM \\
		      \textbf{Display resolution}  & 1366x768 & 1280x1024 \\
             \bottomrule[1pt]
		    \end{tabular}  
	  \end{threeparttable}
	}
  \label{tab:testparameters}
  \vspace{-0.3cm}
\end{table}

\subsubsection{Results}
\label{section:results}              
\emph{Google} sent a security alert via email when either of the features \textit{IP address}, \textit{user agent} or \textit{resolution} changed (see Table~\ref{tab:experimenttwo-google}). Changes in one of the features \textit{language} and \textit{time} didn't result in a warning instead. In contrast to that, we have seen before that strong variations of the IP address result in a request for additional authentication factors (see Table~\ref{tab:ergebnisseVersuch1Teil1}). When modifying two features, all combinations resulted in a security warning, except for the combination of \textit{language} and \textit{time}. Modifying three features resulted at least in a security warning, and the combination of \textit{IP address}, \textit{user agent}, and \textit{time parameters} led to an additional authentication factor requested. Concluding all results, our derived Google feature set contains \textit{IP address} (highest weighting), \textit{time parameters} (lower weighted than IP), \textit{user agent} and \textit{resolution}.

\emph{LinkedIn}'s RBA was triggered with combinations of \textit{IP address} and at least one of the other parameters (see Table~\ref{tab:experimenttwo-linkedin}). Thus, LinkedIn's feature set comprises \textit{IP address}, \textit{user agent}, \textit{language}, \textit{time parameters} and \textit{resolution}. The IP address seems to be higher weighted since it triggered RBA in the prior study alone.

\begin{table}[t]
		\caption{Results of Study 2 for Google modifying a \emph{single feature} (left), \emph{two features} (middle), and \emph{more than two features} (right).\\
		(A: Additional authentication factors requested,
		- : No RBA triggered,
		S: Security alert,
		C: Critical security alert)
		}

	\centering
		\definecolor{hellgrau}{gray}{0.92}
		\makeatletter
		\newcommand*{\minuscellcolor}{}
		\def\minuscellcolor\ignorespaces{%
			\@ifnextchar X{\cellcolor{hellgrau}}{}%
		}
		\newcolumntype{C}{>{\minuscellcolor}c}
	\resizebox{0.7\linewidth}{!}{
					\begin{tabular}{l || c}
						& Result\\
						\hline
						\textbf{IP address}         &  S\\
						\textbf{User agent} &  S\\
						\textbf{Language}   &  -\\
						\textbf{Time}       &  -\\
						\textbf{Resolution} &  S\\
						\hline
					\end{tabular}%
		\hspace{0.5cm}
					\begin{tabular}{l||l|l|l|l|l}
						& \textbf{IP} & \textbf{UA} & \textbf{L} & \textbf{T} & \textbf{R}\\
						\hline
						\textbf{IP address} &      & S   & S   & S   & S \\
						\textbf{User agent} & S    &     & S    & S    & S \\
						\textbf{Language} 	& S    & S   &      & -    & S \\
						\textbf{Time} 		& S    & S   & -    &      & S \\
						\textbf{Resolution} & S    & S   & S    & S    &  \\
						\hline
					\end{tabular}
		\hspace{0.5cm}

						\begin{tabular}{C|C|C|C|C||c}
							\textbf{IP} & \textbf{UA} & \textbf{L} & \textbf{T} & \textbf{R} & Result \\
							\hline
							X&&X&X&&S\\
							&X&X&X&&S\\
							&&X&X&X&S\\
							X&X&&X&&A/C\\
							X&X&X&X&&A/C\\
							X&X&&X&X&A/C\\
							X&X&X&X&X&A/C\\
							\hline
						\end{tabular}
}
		\label{tab:experimenttwo-google}

\bigskip
		\caption{Results of Study 2 for LinkedIn modifying a \emph{single feature} (left) and \emph{two features} (right).\\
	    (A: Additional authentication factors requested,
		- : No RBA triggered)
}
\centering
\resizebox{0.5\linewidth}{!}{
						\begin{tabular}{l || c}
							& Result\\
							\hline
							\textbf{IP address}         &  -\\
							\textbf{User agent} &  -\\
							\textbf{Language}   &  -\\
							\textbf{Time}       &  -\\
							\textbf{Resolution} &  -\\
							\hline
						\end{tabular}%
		\hspace{0.5cm}
					\begin{tabular}{l||l|l|l|l|l}
						& \textbf{IP} & \textbf{UA} & \textbf{L} & \textbf{T} & \textbf{R}\\
						\hline
						\textbf{IP address} &      & A   & A   & A   & A \\
						\textbf{User agent} & A   &      & -    & -    & - \\
						\textbf{Language} & A   & -    &      & -    & - \\
						\textbf{Time} & A   & -    & -    &      & - \\
						\textbf{Resolution} & A   & -    & -    & -    &  \\
						\hline
					\end{tabular}
}
		\label{tab:experimenttwo-linkedin}%
\end{table}

\emph{Facebook} seems to have RBA deactivated by default. We could not trigger RBA on accounts having at least 50 connections to other accounts (friends). However, we could trigger RBA on two female accounts having both 40-50 friends and a high interaction rate based on received friendship requests and messages from other users. Due to the possible dissimilarities between the test accounts (RBA enabled or disabled), we cannot deduce the exact feature set here. However, our results show that Facebook requested additional authentication factors when at least \textit{IP address}, \textit{user agent} and \textit{resolution} were changed.

On \emph{Amazon} and \emph{GOG.com} we could not trigger RBA with more or other parameters than the IP address. Thus, their derived feature sets contain only the \textit{IP address} of our probed features.

The remaining online services \emph{Steam}, \emph{Twitch} and \emph{iCloud} did not show any reaction in both studies. Possible reasons for this behavior could be:
\begin{inparaenum}[(i)]
	\item~RBA was not implemented or not activated by the user behavior.
	\item~Other features than the five tested were rated as more important.
	\item~An internal warning was triggered informing operational staff about suspicious behavior.%
\end{inparaenum}

\subsection{Study 3: Analyzing Additional Authentication Factors}

\begin{table}
		\captionof{table}{Captured additional authentication factors\\
		(*: Authentication factor was offered in all tested parameter variations)}
		\centering
		\small
		\resizebox{0.8\linewidth}{!}{%
				\begin{tabular}{@{}ll@{}}
				    \toprule[1pt]
					\textbf{Service} & \textbf{Requested authentication factors} \\ 
					\midrule 
					Facebook & \makecell[l]{Approve login on another computer*\\Identify photos of friends*\\Asking friends for help*\\Verification code (text message)} \\ 
					\midrule 
					Google & \makecell[l]{Enter the city you usually sign in from\\Verification code (email, text message, app, phone call)\\Press confirmation button on second device (tablet, smartphone)} \\ 
					\midrule 
					LinkedIn & Verification code (email)* \\ 
					\midrule 
					Amazon & Verification code (email*, text message) \\ 
					\midrule 
					GOG.com & Verification code (email)* \\ 
					\bottomrule[1pt]
				\end{tabular}
				
		}
		\label{tab:auth-factors}
	\vspace{-1.5em}
\end{table}

\label{subsection:study-3}
With RBA being triggered, additional authentication factors are requested by the respective online service. Depending on internal account settings, online services might vary the set of requested additional authentication factors. Overviews of neither the additional authentication factors nor the corresponding RBA user interfaces in current practice were published in literature to date. For this reason, we tried to capture as many variations as possible. In order to achieve this, we added a mobile phone number, a smartphone or tablet as a second device and did additional user actions (e.g. writing a private message with phone number included). We triggered RBA on desktop and mobile devices with all possible combinations and monitored the demanded authentication factors (see Table~\ref{tab:auth-factors}).

\section{Discussion}
\label{section:discussion}
According to our findings, all tested RBA-instrumented online services used the \textit{IP address} in their feature sets. Most online services also used additional features as \textit{user agent} or \textit{display resolution}. All tested online services offered verification codes as an additional authentication factor. The test results confirmed our hypothesis that online services rated the \textit{IP address} higher than other parameters.

Facebook's verification code feature leaked the full phone number%
. We consider this as a bad practice and a threat for privacy. In so doing, phone numbers of users can be obtained. Also, attackers can call the number and gain access to the verification code by social engineering. We are convinced that such a RBA solution will not mitigate incentives for credential stuffing or online password guessing attacks. Thanks to the prompt reaction by Facebook, this vulnerability is now fixed: We contacted Facebook about the phone number leak on September 4th, 2018. Facebook resolved the issue on September 6th, 2018. Since this issue seemingly remained undiscovered by Facebook before our disclosure, this underlines the demand for more research on RBA to improve its overall security.

\subsection{Derived RBA Models Applied in Practice}
\label{section:rba_models}
Based on our findings, we are able to derive three distinct types of conceptual RBA models. Note that due to the abstract nature of these models, they do not provide implementation details.

The \textbf{Single-Feature Model} relies on a single feature only. The password authentication process is extended to search for an exact match of the IP address in the IP address history of the user. If there is no such match, additional authentication steps are requested. We assume that GOG.com adopted this model.
This model is easy to implement, since only one feature has to be stored and evaluated. Thus, a minimum of sensitive data has to be collected and stored. However, this approach entails potential usability problems. Since IP addresses might change frequently in time~\cite{alaca_device_2016}, this can result in frequent re-authentication. Hence, we do not consider this as a sensible RBA solution for practical use.

The \textbf{Multi-Features Model} extends the single-feature model. It derives additional features from the IP address. These are evaluated together with additional features in a scoring model, which compares the current feature values with the authentication history. Depending on the resulting risk score, multiple types of actions are performed (e.g. sending security alerts or requesting additional authentication factors). According to our observations this model was adopted by Google and---in slightly more simplified form without security alerts---by Amazon and LinkedIn.
This model has the potential to increase usability compared to the single-feature model since additional authentication factors can be requested less frequently. However, attackers are possibly able to learn about the RBA implementation based on detailed information delivered in security alerts.

The \textbf{VIP Model} protects only special users.
Depending on the user's status (e.g. important or not important), RBA is active or inactive. We assume that Facebook used this model.
This procedure will make it harder for attackers to gain information about the used RBA implementation. However, if such a mechanism is known, attackers are able to find out whether an account is considered as important by the online service (which is the case when RBA is triggered). Also, this model puts some users at risk since it does not protect all users.

\vspace{-0.5em}
\subsection{Limitations}
\label{section:limitations}
We were able obtain a high amount of information with the described studies. However, the RBA behavior could only be determined from visible reactions disclosed by the online services. Hence, we can only estimate internal weightings for features. It is still possible that the real weightings might vary in detail. In addition, RBA is required to be activated anytime for determining feature sets accurately. It is still possible that online services (additionally) use other features which were not tested in the studies (e.g. canvas fingerprinting).
	
Although we took a lot of care of not being detectable as an automated user, we cannot fully exclude that the inspected online services identified our identities as non-humans. Judging some of the hints we obtained during our pilot phase, we are strongly convinced, though, that our investigations remained under respective detecting thresholds.

\subsection{Ethical Considerations}
\label{section:ethics}
It is commonly found that tools and techniques used for security analysis are ``dual use'', i.e., can be used for illegitimate purposes as well. We believe our work is justified, as the expected security gain (from broader adoption of RBA) outweighs the expected security implications. Furthermore, we designed our study to keep the potential impact on the server infrastructure minimal. Finally, we followed the principle of responsible disclosure%
.

\section{Conclusion}
\label{section:conclusion}
RBA is becoming more and more important to strengthen password-based authentication without affecting the user interface at the same time. As RBA is still in its infancy, it is of paramount importance that RBA approaches and implementations are rigorously analyzed following common scientific policies. Unfortunately, almost all early adopters of RBA restrain their approaches and experiences, preventing the required scientific dialogue and the widespread adoption. To close this information gap, we developed distinct studies enabling to verify whether a particular online service adopted RBA. Moreover, we were able to determine the underlying feature sets and requested authentication factors%
.

We can confirm the general trend in RBA of using the IP address as a high weighted indicator to determine risks of login attempts. Some services also used additional lower weighted indicators (e.g. user agent). Furthermore, verification codes are currently the unwritten standard for additional RBA authentication factors%
. Our research disclosed potential vulnerabilities and usability problems on specific RBA implementations (one vulnerability was fixed after we contacted the company in charge). Since RBA usually evaluates sensitive data, there is need for more open research on this technology to mitigate such potential risks.

\vspace{-1em}
\subsubsection*{Acknowledgement}

This research was supported by the research training group ``Human Centered Systems Security'' (NERD.NRW) sponsored by the state of North-Rhine Westphalia.

	\bibliographystyle{splncs04}
	\bibliography{main}

\end{document}